# Understanding oxide-thickness-dependent variability in dense Si-MOS quantum dot arrays

A. Loenders[1,2]†, J. Van Damme[2]*†, C. Godfrin[2], P. Favia[2], J. Franco[2], T. Van Caekenberghe[1,2], B. Raes[2], G. Jaliel[2], S. Baudot[2], L. F. Pinotti[2], A. Grill[2], G. Simion[2], K. Moors[2], V. Levajac[3,2], S. Beyne[2], S. Sharma[2], S. Kubicek[2], Y. Shimura[2], R. Loo[2,4], M. Mongillo[2], D. Wan[2], K. De Greve[1,2,5]

[1]Department of Electrical Engineering (ESAT), KU Leuven; Leuven, B-3000, Belgium.
[2]Imec; Leuven, B-3001, Belgium.
[3]Department of Physics, KU Leuven; Leuven, B-3000, Belgium.
[4]Department of Solid-State Sciences, Ghent University; Ghent, B-9000, Belgium.
[5]Proximus Chair in Quantum Science and Technology, KU Leuven; Leuven, B-3000 Belgium.

*Corresponding author. Email: jacques.vandamme@imec.be

†These authors contributed equally to this work.

**Abstract**: Achieving uniform and scalable control of semiconductor spin qubits remains a key challenge for large-scale quantum computing. In this work, we investigate how gate-oxide thickness influences uniformity in dense two-dimensional silicon quantum dot arrays. Using a 7 × 7 array fabricated in a 300 mm CMOS-process patterned by EUV lithography, we statistically characterize 392 quantum dots across four different oxide thicknesses. The threshold voltages, capacitances, lever arms, and charging energies are extracted using parallel row-based measurements and we identify an optimal $SiO_2$ thickness of 17 nm that minimizes threshold-voltage variability below 63 mV standard deviation. Our observations illustrate how multiple sources of disorder can introduce competing oxide-thickness dependencies, resulting in non-monotonic trends. These results provide key design guidelines for dense, scalable silicon spin-qubit architectures.

## I. INTRODUCTION

Scaling quantum computers beyond tens of thousands of physical qubits toward practical applications [1] remains a formidable challenge across all currently pursued technology platforms [2–4]. Solid-state implementations such as superconducting qubits and spin qubits in quantum dots offer a key advantage: they can be fabricated with industrial processes developed for complementary metal-oxide-semiconductor (CMOS) transistors [5,6], and they can be co-integrated with classical electronics to mitigate wiring bottlenecks [7–10]. Two-dimensional grids of spin qubits in quantum dots can, in principle, be realized with unit-cell footprints that are six orders of magnitude smaller than those of state-of-the-art superconducting qubits. This density would allow millions of qubits to fit on a single die, in contrast to at most a few tens of thousands of superconducting qubits on an entire wafer [11], thereby alleviating the need for complex quantum interconnects between dies, wafers, or cryostats [12,13].

Scaling spin qubits in gate-defined quantum dots to larger two-dimensional arrays is, however, particularly challenging in the presence of local disorder. Spurious dots arising from strain [14–17], charge defects [18], interface roughness [19], or work function variations [20] complicate the simultaneous tuning of dots into the desired few-electron occupation regime and prevent achieving the intended exchange coupling between neighboring quantum dots [21]. Enhancing dot uniformity in dense arrays would directly improve tune-up procedures and ultimately enable reduced fan-out and wiring through common-gate architectures [22,23]. Assessing such uniformity requires significant device statistics, which are difficult to obtain with traditional measurement setups that are limited by the number of available control lines and demanding packaging constraints. Recent progress, including cryogenic wafer-scale probers [24], coupon-scale probing [25], and multiplexed grids of small devices [26,27] has opened the door to systematic uniformity studies. However, uniformity within dense two-dimensional arrays of dots has not yet been investigated. In this work, a dense two-dimensional 7 x 7 quantum dot array featuring row-wise, seven-fold parallel measurement capability is introduced (Fig. 1 A). The number of required lines ($T$) for measurements as a function of dots ($D$) scales more favorably in this dense $N$ x $N$ 2D architecture ($T \sim \sqrt{D}$) compared to probing dots in isolated samples ($T \sim D$), following Rent's rule [28,29]. This test-device enables an evaluation of various aspects of quantum dot uniformity in our overlapping-gate silicon MOS spin qubit platform, previously reported in Ref. [30,31]. In particular, we identify the gate-oxide thickness that optimizes quantum dot uniformity.

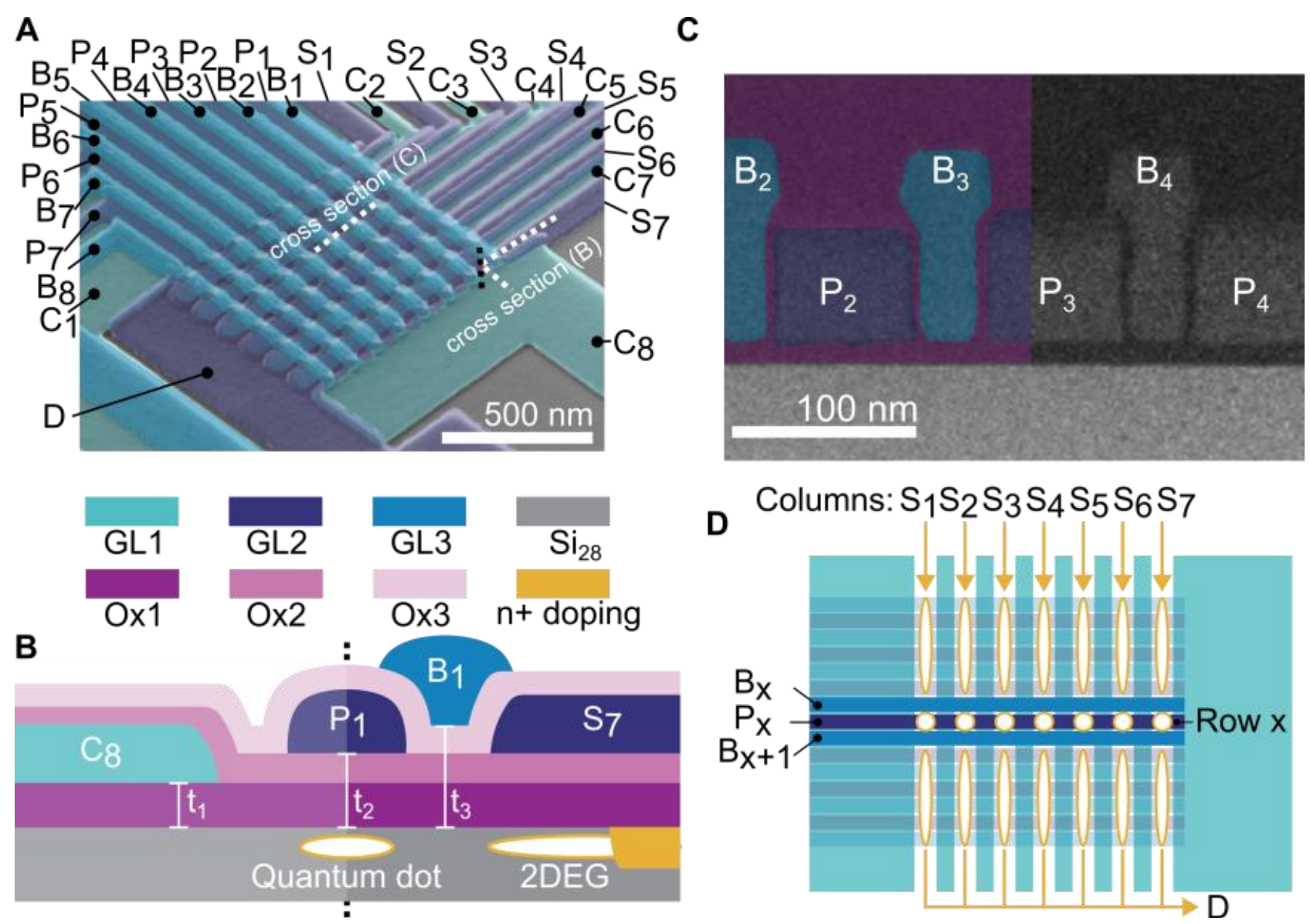

**Fig. 1. Overlapping-gates SiMOS 7x7-array of quantum dots.** (**A**) False colored scanning electron microscope (SEM) image with a color-code representing the three different gate layers. The confinement gates are patterned in the first gate-layer (GL1) and labeled as $C_1$, …, $C_8$, forming the columns of the sample. The plungers are patterned in the second gate-layer (GL2) and labeled as $P_1$, …, $P_7$, each controlling a row of seven quantum dots. The accumulation gates are also patterned in GL2 and bring the two-dimensional electron gas (2DEG) from the drain and seven source implant regions to the array columns. They are labeled as D, $S_1$, …, $S_7$ respectively. The barrier gates separating the rows of quantum dots are defined in the third gate-layer (GL3) and labeled as $B_1$, …, $B_8$. (**B**) Schematic of the angled cross-section along the cut

shown in (A), with $t_1$, $t_2$, $t_3$ the net oxide thickness for the different gate-layers. (**C**) False-colored HAADF-STEM image of a cross-section at the location shown in (A) for a sample with 8 nm $t_1$ oxide thickness. (**D**) Parallel measurement of seven quantum dots formed under $P_x$, with $B_x$ and $B_{x+1}$ acting as source and drain barriers respectively. The confinement gates form seven current paths (yellow arrows), allowing for parallel measurement of transport through the quantum dots.

## II. EXPERIMENT DESCRIPTION

The 7 x 7 two-dimensional quantum-dot array comprises forty-nine quantum dots with a 110 nm dot-to-dot pitch and is fabricated using 300-mm-wafer SiMOS technology developed at Imec. The difference with the previous demonstration is that all electron-beam lithography steps are replaced with EUV lithography [32]. The sample employs an overlapping gate architecture in which three highly doped polysilicon gate-layers define the accumulation, confinement, plunger, and barrier potentials required to form and operate the quantum dots (Fig. 1 A). Each gate-layer is deposited and patterned sequentially, separated by $SiO_2$ dielectrics. The first gate-layer oxide is thermally grown, while for the second and third gate-layer oxides, a high-temperature chemical vapor deposition process is used at 780 degrees Celsius. Consequently, the gates in the first, second, and third layers are separated from the enriched $Si_{28}$-substrate by different oxide layers with net thicknesses $t_1$, $t_2$ and $t_3$, as illustrated in the schematic cross-section of Fig. 1 B.

The quality of the $SiO_2$ gate dielectrics is a key determinant of overall qubit performance, as charged defects within the oxides act as sources of noise and spurious dots [33–35]. Additionally, differences in thermal-expansion coefficients of the oxide and the gate materials generate strain at the interfaces, producing local modulation of the conduction-band edge and contributing further to disorder and spurious dot formation [14,15,36]. This strain and sensitivity to defects can be mitigated by letting the oxide relax over a longer distance and by increasing the distance between the gate and defects, therefore the thickness of the primary oxide layer plays a significant role in defining and mitigating the disorder landscape experienced by the quantum dots. In general, a non-trivial interplay is expected between strain, charged-defect density, gate lever arms, and proximity of higher-stack oxides with potentially lower material quality, all of which depend sensitively on the primary oxide thickness. Therefore, we are performing a detailed study on the impact of this primary oxide thickness $t_1$ on dot uniformity.

**Table 1. Sample overview.** Eight samples, taken from four wafers, with different net deposited oxide thickness. The die coordinate specifies the location on the wafer (with respect to the center) and the subdie number (out of 15) specifies the chip on that die. Oxide thicknesses $t_1$, $t_2$, and $t_3$ correspond to the nominal net deposited oxide thickness underneath each gate layer as illustrated in schematic in Fig. 1 B.

| Sample name | Wafer | Die coordinate (X,Y) | Subdie | $t_1$ (nm) | $t_2$ (nm) | $t_3$ (nm) |
|---|---|---|---|---|---|---|
| A | Wafer 1 | (0,1) | 14 | 8.0 | 13.0 | 18.0 |
| B | Wafer 1 | (1,1) | 15 | 8.0 | 13.0 | 18.0 |
| C | Wafer 2 | (-1,1) | 15 | 12.0 | 17.0 | 22.0 |
| D | Wafer 2 | (2,2) | 15 | 12.0 | 17.0 | 22.0 |
| E | Wafer 3 | (-2,2) | 14 | 15.0 | 20.0 | 25.0 |
| F | Wafer 3 | (-2,2) | 15 | 15.0 | 20.0 | 25.0 |
| G | Wafer 4 | (1,2) | 14 | 20.0 | 25.0 | 30.0 |
| H | Wafer 4 | (-1,2) | 14 | 20.0 | 25.0 | 30.0 |

Eight samples spanning a range $t_1$ values were compared (see Table 1), with all other process parameters held nominally the same. After the deposition of each gate layer, the sample undergoes an etching process that slightly over-etches into the underlying oxide layer, reducing the effective oxide thickness from the expected values in Table 1. From the false-colored high-angle annular dark-field scanning transmission electron microscopy image (HAADF-STEM) for a sample with 8 nm $t_1$ oxide (Fig. 1 C), it is possible to extract the effective thicknesses $t_1$, $t_2$ and $t_3$ after processing (see Appendix). The oxide thickness $\delta_2 = t_2 - t_1$ is found to be ~4.6 nm, while $\delta_3 = t_3 - t_2$ is ~0.8 nm due to a longer etching process. Since the deposition and etching process does not change for the eight samples measured, these inter-gate-layer oxide thicknesses are taken as approximations for all samples. For each sample, disorder-sensitive metrics like gate threshold voltages, dot capacitances, lever arms, and charging energies were extracted in a dense 2D-array environment [37]. For every oxide-thickness condition, two times forty-nine (98) quantum dots (392 dots across all samples) were measured. This ensures sufficient statistics to quantify the spread of these metrics and assess the impact of the primary oxide thickness on sample uniformity.

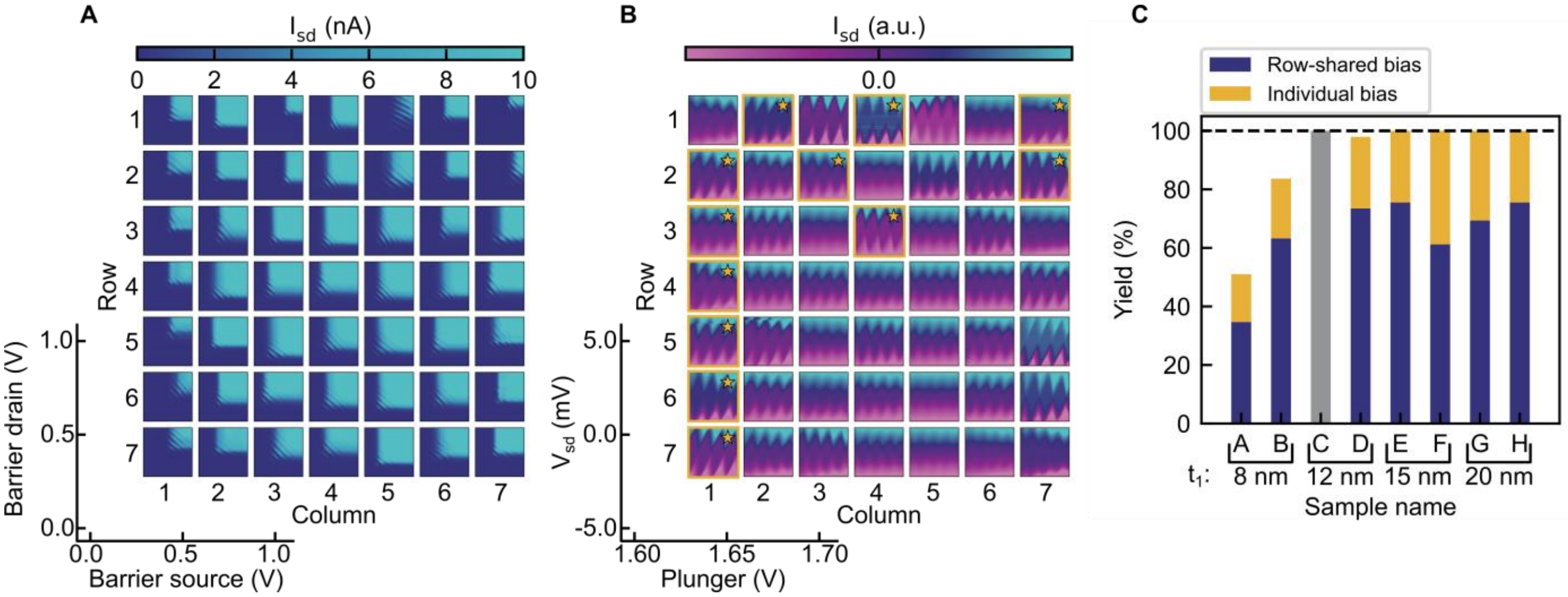

**Fig. 2. Quantum dot yield.** (**A**) Barrier-maps of all 49 dots in sample E, captured with seven sequential measurements, row-by-row, using seven parallel measurement channels on sources $S_1$, ..., $S_7$, as explained in the text. (**B**) Coulomb diamonds of all 49 dots in sample E, captured with row-shared barrier-bias voltages based on the barrier-maps in (A), see text. The Coulomb diamonds highlighted with yellow stars were taken with individual barrier-bias values. (**C**) Yield overview of the eight measured 7 x 7-arrays (four pairs of different oxide thickness $t_1$). The blue bars represent the fraction of dots in the array where the Coulomb diamonds could be extracted under a common row barrier-bias, while the yellow bars correspond to the yellow-star-highlighted

diamonds in (B) that required individual biasing conditions. Sample C was measured under a different wiring scheme without parallel measurement capabilities (visualized as a grey bar), where all dots were biased and measured individually.

To measure each of the 7 x 7 arrays, the source and drain accumulation gates ($S_1$-$S_7$ and D) are positively biased to form a 2D-electron gas (2DEG) that connects the implant regions to the active sample area. A negative bias is applied to the confinement gates ($C_1$-$C_8$) to define seven columns that act as separate current channels pinching off the rest of the 2DEG. For each measurement, the desired row for measurements is subsequently selected by configuring the biases such that gate $P_x$ acts as the plunger for the quantum dots in that row, while the gates $B_x$ and $B_{x+1}$ serve as the source-barrier ($B_s$) and drain-barrier ($B_d$), respectively. All other plunger and barrier gates are biased positively to extend the 2DEG from the source and drain accumulation regions to the row-under-test, see Fig. 1 D. By using seven parallel current sensors located at the sources, it is possible to measure the transport current through the seven quantum dot channels simultaneously, significantly reducing the number of measurements required to characterize the full array.

First, the $P_x$, $B_s$, and $B_d$ gate voltages are swept individually to determine the turn-on voltages of each dot. Based on these measurements, a 2D-barrier map (Fig. 2 A) is acquired by sweeping the $B_s$ and $B_d$ voltages whilst holding the plunger voltage fixed above the threshold. These measurements are performed using the parallel-row method, simultaneously extracting all seven barrier maps with a single measurement.

From the barrier maps, a common barrier bias point is chosen for the entire row such that most dots exhibit Coulomb oscillations and allow for the measurement of Coulomb diamonds (Fig 2 B). However, in some cases no single bias point produces Coulomb diamonds for all seven dots. In those instances, the affected dots are measured using locally optimized barrier-bias values, requiring an additional measurement. These Coulomb diamonds are marked with a yellow star in Fig. 2 B.

## III. RESULTS

### A. Dot Yield

Combining barrier maps and Coulomb diamonds data enables the definition of two yield metrics for the 7 x 7 array (Fig. 2 C). The row-shared Coulomb diamond yield represents the fraction of dots for which Coulomb diamonds could be measured under row-shared bias conditions, while the total yield also includes the diamonds that were biased individually. The divergence from 100% yield in sample D is due to human error, causing only 48 of the 49 diamonds to be properly measured. In sample A, columns six and seven could not be used due to a short between $S_6$, $S_7$, and $C_8$, while sample B did not achieve turn-on of column seven (see Appendix). Sample C was measured differently due to a different wiring scheme without parallel measurement capabilities, and thus all dots were measured with an individual bias. The dots located at the edges of the array consistently display different operating windows compared to those in the center (Fig. A6), likely due to the geometric difference of the confinement gates at the boundary.

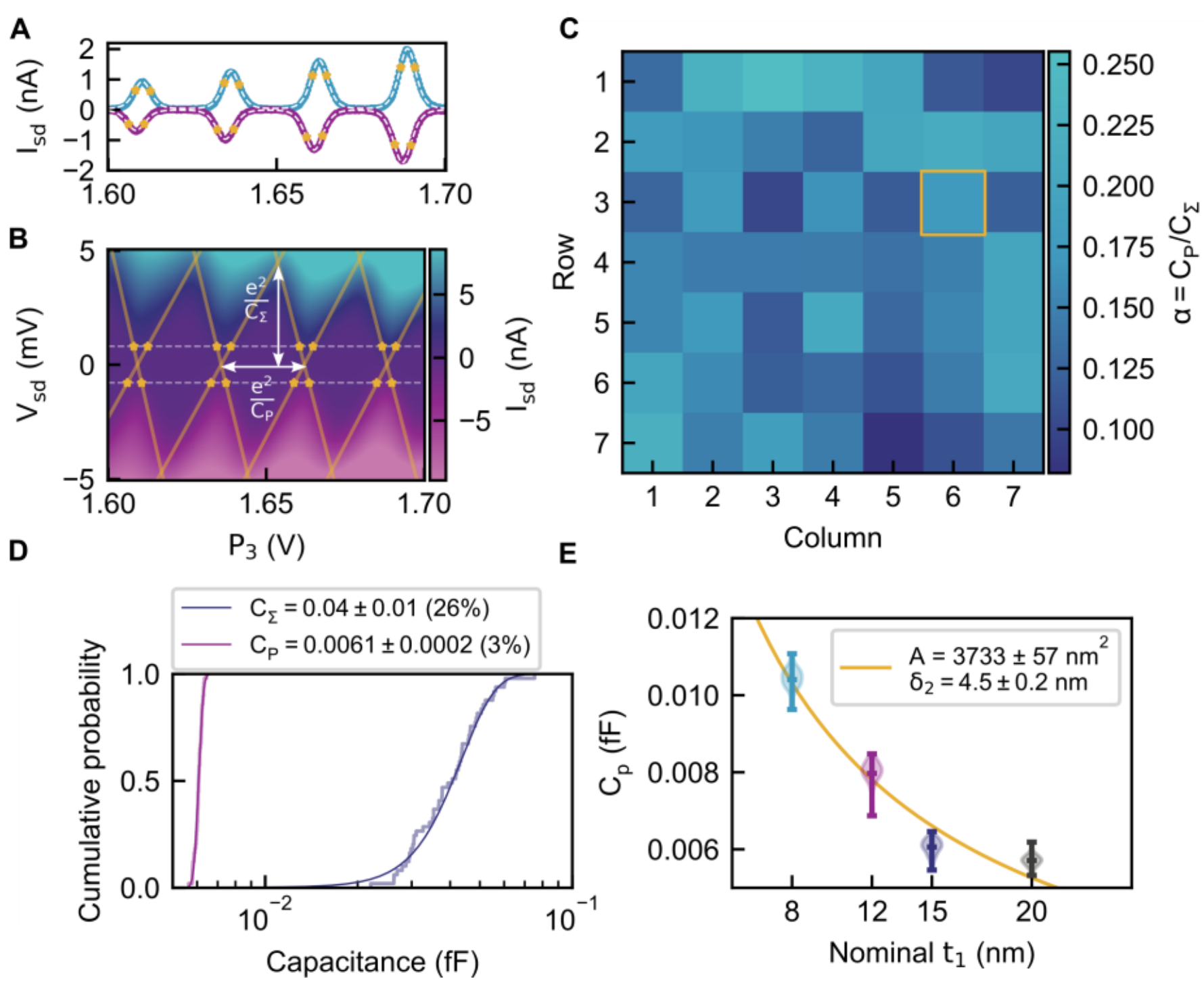

**Fig. 3. Coulomb diamond uniformity.** (**A**, **B**) Example Coulomb diamonds of dot (3,6) of sample E. The diamond width and height are extracted from linear fits (orange lines) through the maximal slopes of the Coulomb peaks at the horizontal line-cuts shown in (A) and relate to the plunger capacitance $C_P$ and total dot-capacitance $C_\Sigma$ respectively. (**C**) Extracted lever-arms from the fitted diamonds at each dot location of sample E. (**D**) Estimated cumulative distribution functions of the total dot-capacitance and plunger capacitance of sample E, with Gaussian fits. The legend shows their respective mean value, standard deviation and relative standard deviation. (**E**) Violin plots of the plunger capacitance distributions of all eight samples grouped by nominal oxide thickness $t_1$. The solid line is a fit to a parallel-plate capacitor model with the dot area A, and the inter-gate-layer oxide thickness $\delta_2$, as free fitting parameters.

## B. Dot capacitance estimate

From the Coulomb diamond measurements, the lever arm α, charging energy $E_C$, plunger capacitance $C_P$ and total dot capacitance $C_\Sigma$ are extracted [37]. For each quantum dot, two line-cuts are taken near zero source-drain bias (Fig. 3 A). The points of maximal slope are identified as the diamond edges (Fig. 3 B), and from the fitted diamond (Fig. 3 B), the width and height directly yield $C_P$, $C_\Sigma$ and $E_C$, while the ratio of the capacitances provides the lever arm α (Fig. 3 C).

The spatial distribution of the lever arms and charging energies across the array shows no evident position-dependent trend (Fig. 3 C, Fig. A7 and Fig. A8), however, a clear difference in variability is observed between the total and plunger capacitances within the arrays (Fig. 3 D). The cumulative probability distribution of each reveal that while the total capacitance exhibits a relative standard deviation spread of 26%, the plunger capacitance varies only by 3% for sample E. This behavior reflects the physical origin of each capacitance, as the total capacitance includes contributions of all electrostatic couplings to the dot, including parasitic capacitances to neighboring gates and dots in the dense array, and the overlap of the electron wavefunction to the source and drain reservoirs. The plunger capacitance, however, is defined only by the quantum dot size and $t_2$ oxide thickness. By modeling the plunger capacitance as a parallel-plate capacitance to the first order, both the quantum dot area and oxide thickness $\delta_2$ (Fig. 3 E) are estimated. The extracted dot area of 3733 $nm^2$ and oxide thickness between GL1 and GL2 $\delta_2$ = 4.5 nm agree well with the expected values of 3500 $nm^2$, corresponding to 50 x 70 nm elongated dots after processing (see Appendix), and 5 nm deposited oxide after a slight GL1 over-etch (see Appendix), respectively. This close agreement indicates precise control of both the sample geometry and dielectric thicknesses during fabrication.

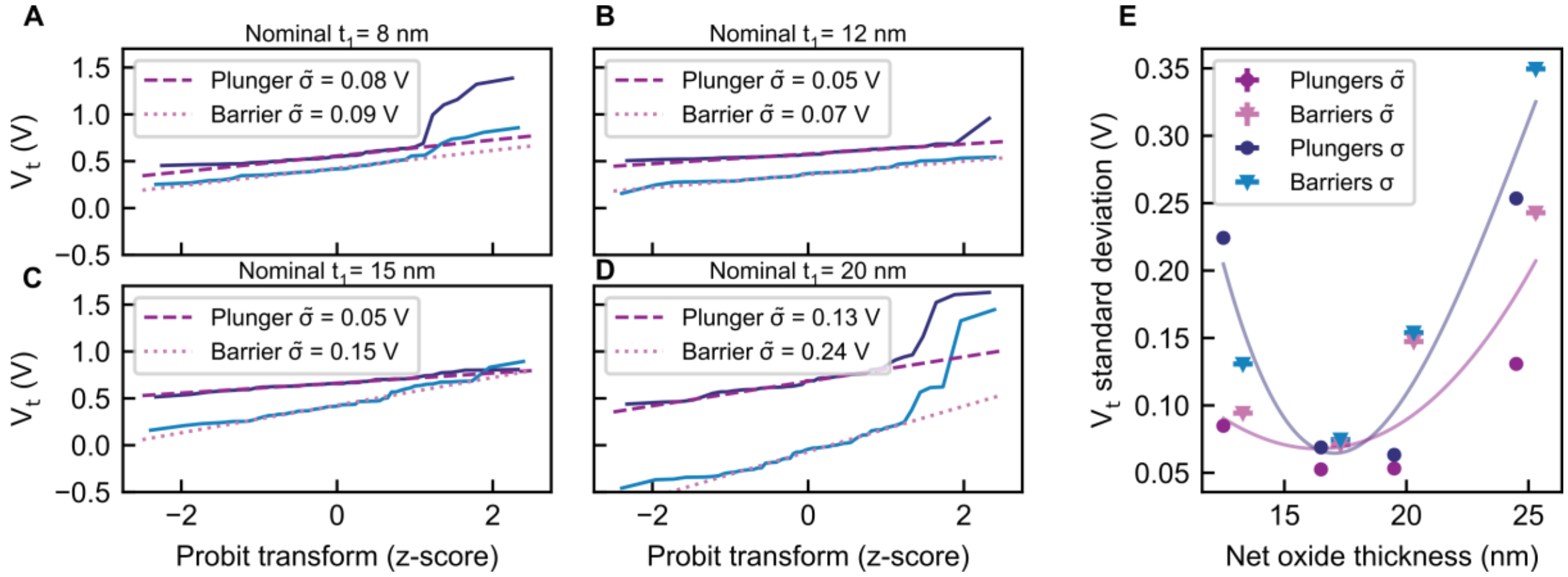

**Fig. 4. Threshold voltage uniformity of central 5 x 5-array.** (**A**, **B**, **C**, **D**) Threshold voltage distributions of all plunger gates (dark blue lines) and all barrier gates (light blue lines), grouped by nominal oxide thickness $t_1$ of the sample. The values are sorted and plotted over the probit transformed x-axis. Linear fits (dark purple for plunger data, and pink for barrier data) to the central region (z-score ∈ [-1;1]) extract the standard deviations of the Gaussian component of the distributions, as explained in the text. (**E**) Threshold voltage standard deviations of the entire distributions ($\sigma$) and Gaussian estimations ($\tilde{\sigma}$) of plunger and barrier gates in (A, B, C, D) plotted as function of their respective oxide thickness, $t_2$ and $t_3$ extracted from Figure 3 (E) and Figure S1. Solid lines are smoothing splines as guides to the eye.

### C. Threshold-voltage variability

From the turn-on I-V characteristics of the plunger and barrier gates, the threshold voltage $V_t$ can be extracted. A sigmoid function is fitted to each I-V curve, and the threshold voltage is defined as the voltage corresponding to the maximum slope of the fit (see Appendix). Using this procedure, threshold voltage distributions of the central 5 x 5 sub-array are obtained for all samples, hereby excluding the outer rows and columns with asymmetric neighboring gates. The standard deviation ($\sigma$) is calculated for each oxide thickness (Fig. 4 A-D).

To suppress the influence of large statistical outliers, the extracted threshold voltages are plotted using a probit-transform [38,39]. First, the extracted threshold-voltage values are sorted and assigned ranks from 1 to *N*, where *N* is the total number of data points. From the ranked dataset, the empirical cumulative distribution function (CDF) is computed, and each value is transformed using the inverse CDF of the standard normal distribution. This effectively maps the data onto the respective percentiles under an assumed Gaussian distribution, with z-scores corresponding to the percentile the data is in, centered around zero.

A z-score filter is then applied where we retain only data points with $z \in [-1;1]$, which corresponds to the central 68 % of an approximately Gaussian-distributed dataset. This procedure allows a Gaussian fit to be performed that ignores strong outliers (Fig. 4 A-D). The resulting standard deviations, with ($\tilde{\sigma}$) and without ($\sigma$) probit-filtering, are plotted as a function of the actual oxide thickness (Fig. 4 E). The actual oxide thickness is determined using the nominal wafer $t_1$ value combined with the inter-gate-layer oxide thickness $\delta_2$ (4.5 nm) between GL1 and GL2 estimated from the capacitance analysis in the previous section (in agreement with the data extracted from the TEM cross-sectional image in Fig. A3). The inter-gate-layer oxide thickness $\delta_3$ between GL2 and GL3 (0.8 nm) is taken from the TEM cross-sectional image in Fig. A3.

For the plunger gates, the standard deviation $\sigma$ (filtered $\tilde{\sigma}$) ranges from 63 (52) mV to 253 (131) mV, while for the barrier gates the standard deviation $\sigma$ (filtered $\tilde{\sigma}$) ranges from 74 (70) mV to 350 (242) mV. The minimum plunger variability occurs at a $t_1$ oxide thickness of 15 nm, corresponding to a $t_2$ thickness of 19.5 nm. For the barrier gates, the minimum is obtained for $t_1$ = 12 nm equivalent to $t_3$ = 17.3 nm.

### D. Spurious dots

Spurious dots are quantum dots formed at unintended locations, typically due to defects or strain in the sample [15]. When such a spurious dot forms near an intended quantum dot, the two can become capacitively or tunnel-coupled, effectively creating an unintended double-quantum-dot system. In this configuration, electrons can move between the intended dot and the spurious dot, partially bypassing the transport channel controlled by the source and drain barriers. This reduces the controllability of the intended dot and, in severe cases, can render it completely unusable.

The presence of a spurious dot forming a double-dot system can be identified in the barrier maps of the intended quantum dot. In addition to the expected diagonal Coulomb oscillations, corresponding to electron tunneling between the dot and the source-drain channel, the interdot tunneling between the intended dot and the spurious dot generates additional Coulomb oscillation lines with a different slope or orientation, depending on the location of the spurious dot.

From Fig. 2 A (and Fig. A5), we observe that these spurious oscillations are consistently perpendicular to one of the two barrier gates (see dot (3, 2) from sample B in Fig. A5). This indicates that the corresponding spurious dot couples predominantly to one barrier gate and therefore must be located in close proximity to it. Moreover, spurious dots that depend on the drain barrier of one row subsequently appear as features depending on the source barrier of the row below (see dot (3, 2) and (4, 2) from sample B in Fig. A5). This behavior strongly suggests that the spurious dots form beneath the barrier gate that separates the two rows, consistent with the sample geometry in which the drain of one dot serves as the source of the dot in the row below. We do observer that all spurious dots appear only within one column, suggesting that the channels within each column are fully screened from each other using the confinement gates.

## IV. DISCUSSION

The analysis presented in Fig. 4 E reveals an optimal net gate-oxide thickness of approximately 17 nm, at which the threshold-voltage variability reaches a minimum ($\sigma$ = 63 mV, $\tilde{\sigma}$ = 52 mV). This optimum, however, pertains strictly to threshold-voltage variation and uniform dot biasing, but does not necessarily translate into optimal conditions for other qubit-relevant parameters (valley splitting, charge-noise, etc.) which also depend on the local electrostatic environment, shaped in part by the gate-oxide via the impact on the lever arm. These additional metrics, not investigated here, may shift or even compete with the identified optimum when considering overall qubit performance.

To interpret the data and understand the observations in Fig. 4 E, we consider that two different competing effects are present in the sample: For oxide thicknesses exceeding 17 nm, the increasing threshold-voltage variability can be attributed to a reduced gate-to-channel lever arm, decreasing the ability of the gate to compensate trapped charges and capacitance. This behavior is consistent with classical MOSFET scaling arguments and Pelgrom-type variability, which predict increased threshold-voltage fluctuations with reduced capacitive coupling, with a linear dependence of the threshold-voltage variability on the oxide thickness [40–42]. Assuming a non-zero scaling of the defect density with oxide thickness, thicker oxides also host a greater absolute number of defects, further exacerbating sample-to-sample variation.

In contrast, the increased variability observed for oxide thicknesses below 17 nm clearly has a different origin compared to thicker oxides. One potential explanation involves considering the effect of thermomechanical strain [14–17,43]. At low temperatures, the differential thermal contractions between the gate material and the dielectric gate oxide produces significant strain at the channel-oxide interface, which scales inversely with the primary oxide thickness. This strain locally modulates the conduction-band edge, generating local wells in the electrochemical potential that can trap spurious electrons [15,36]. Such strain-induced spurious dots could distort the transport channel and shift the threshold voltages of the surrounding gates, naturally explaining the increase in variability at the smallest oxide thicknesses, which would be in line with some of our previous observations in isolated samples [36].

The overlapping-gate architecture employed in this study imposes additional constraints on exploiting the identified oxide thickness optimum. Different gate layers reside atop distinct dielectric thicknesses $t_1$, $t_2$, $t_3$ and any attempt to optimize the oxide thickness for one layer inevitably introduces nonuniformity in others, preventing a global optimization strategy across all gates. A single-gate-layer architecture, in which all gates rest on the same oxide thickness, would present a more suitable platform for leveraging an optimized gate-oxide thickness for uniform gates and dots across the entire dense array [44,45].

Finally, the spatial pattern of spurious-dot signatures observed in the barrier maps (Fig. 2 A and Fig. A5) suggests where the dominant variability could originate from. Spurious dots consistently nucleate under the barrier gates, shared between two neighboring rows, but confined within a column. This interpretation aligns with our capacitance analysis, which shows that total dot capacitance variability is not limited by the plunger capacitance, but most likely dominated by coupling to source and drain. The dot electron wave-function overlap with the source and drain regions depends exponentially on the barrier potential and is therefore extremely sensitive to barrier variation.

Together, these observations establish a coherent picture in which oxide thickness, cryogenic strain, and barrier-gate disorder jointly shape the uniformity of gate-defined quantum dots. This identifies the barrier region as the principal locus of variability and highlights barrier-field engineering and materials uniformity as key directions for improving reproducibility in large-scale quantum dot arrays.

**Acknowledgments**
The authors are grateful for the processing support from the Imec P-line, to P. Carolan and A. Impagnatiello for metrology support, and to K. Verhemeldonck and E. Vandenplas for aiding with the preparation of the samples. The authors would also like to thank R. Li, A. Elsayed, and J. Chu for the original 7 x 7 array designs.

**Funding:**
This work was supported in part by European Union's Horizon Europe Program under grant agreement No 101113983 (Qu-Pilot). This work was performed as part of Imec's Industrial Affiliation Program (IIAP) on Quantum Computing.

**Author contributions:**
Conceptualization: AL, JVD, CG
Methodology: AL, JVD
Investigation: AL, JVD, PF, JF, TVC, BR, JG, SBA, LFP, AG, GS, KM, VL, SBE, SS, SK, YS, RL
Visualization: AL, JVD, PF
Funding acquisition: DW, KDG
Project administration: CG, MM, DW, KDG
Supervision: CG, KDG
Writing – original draft: AL, JVD
Writing – review & editing: AL, JVD, CG, PF, JF, TVC, BR, JG, SBA, LFP, AG, GS, KM, VL, SBE, SS, SK, YS, RL, MM, DW, KDG

**Competing interests:**
The authors declare they have no competing interests.

**Data, code, and materials availability:** All data needed to evaluate and reproduce the results shown in the paper are available on Zenodo at https://doi.org/10.5281/zenodo.20133452.

## APPENDIX

### *Measurement setup*

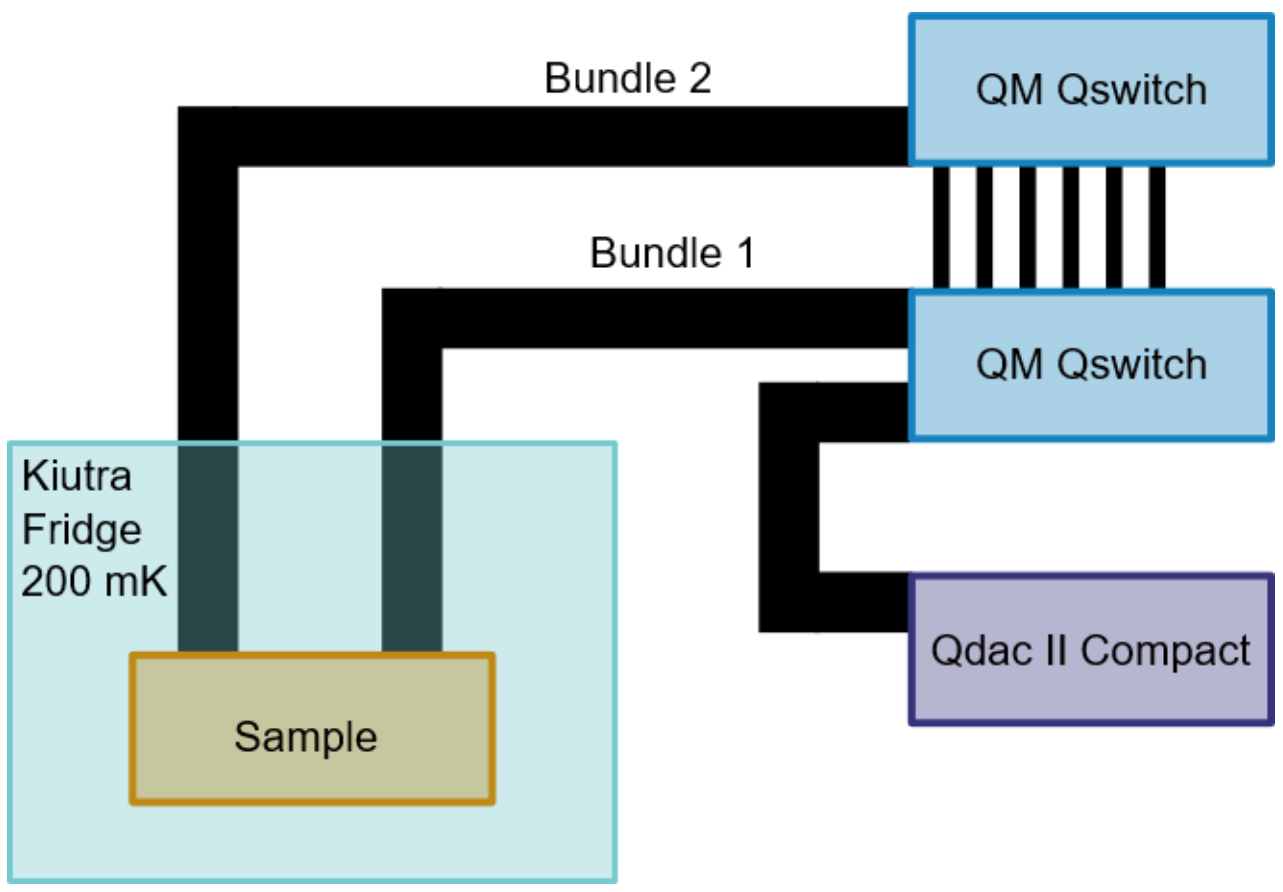

**Fig. A1. Experimental setup.** The Kiutra fridge has a single ADR unit for cooling, allowing the measurements to be performed at 200 mK, the bundles are 24 wire shielded twisted pair Fisher cables connecting the fridge to the QSwitches on the measurement rack. The two Quantum Machines QSwitches each have eight BNC in/output connections, six of which connect the two devices, while the QDac II Compact from Quantum Machines has 24 channels connected to the QSwitch on bundle 1.

The sample is wire-bonded to a custom PCB before being loaded into a Kiutra ADR fridge with a sample temperature of 200 mK (see Appendix), which is connected to the measurement setup via two Fischer multi-line cables (bundle 1 and bundle 2), each providing twenty lines to the sample. On each cable, a QM QSwitch module with eight outputs is installed and on the QSwitch connected to bundle 1, an additional 24-channel QM QDAC II Compact is mounted. This unit supplies all DC gate voltages and measures all source-drain currents in the sample by using all twenty-four channels (Fig. A1).
The two QSwitch modules are interconnected by connecting six of their eight BNC outputs. These connections allow the experimenter to dynamically reassign which fridge lines are routed to the room-temperature electronics. Four of these linked channels use unused outputs of the 24-channel QDAC II Compact, enabling independent control of a plunger gate and the source and drain barriers. The remaining interconnections are routed to multiple plunger, barrier and confinement gates simultaneously, as these do not require individual bias control. For example, a single shared bias is used for all barrier and plunger gates that extend the 2DEG from the source and drain accumulation regions to the row under test.

***Threshold voltage extraction***

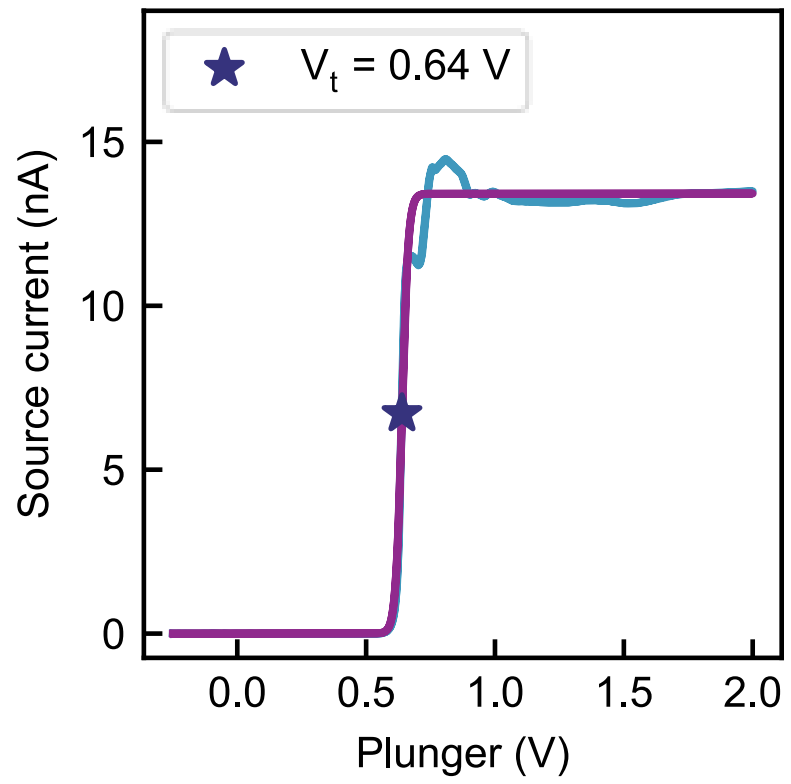

**Fig. A2. Threshold voltage extraction.** In the IV-curve for the plunger of dot (3,5) of sample E is shown in light blue. A sigmoid fit is applied to the data, yielding the purple curve. From the fit, the threshold voltage is extracted and marked as a dark blue star.

To extract the threshold voltage, the I-V curve is fitted using a sigmoid function:

$$f\left(V_{\mathrm{g}}, V_{\mathrm{t}}, k\right) = \frac{I_{\mathrm{max}}}{1 + e^{-k*(V_{\mathrm{g}} - V_{\mathrm{t}})}} \tag{1}$$

where $I_{max}$ is the maximal current, $V_g$ the gate voltage, $V_t$ the extracted threshold voltage and $k$ a free fitting parameter. Applying the sigmoid fitting to the plunger turn-on of dot (3,5) of sample E yields a threshold value of 0.64 V (Fig. A2).

***Transmission electroscope microscopy and energy-dispersive X-ray spectroscopy***

Two cross-sectional images are taken using transmission electroscope microscopy (TEM) and energy-dispersive X-ray spectroscopy (EDS) from an array with a nominal $t_1$ oxide thickness of 8 nm. The first cross-section is taken horizontally through the array, following a plunger gate, see Fig A3 A-C. From this cross-section it is possible to determine $t_1$ as ~9 nm and $t_2$ ~13,6 nm. The second cross-section in Fig A3 D-F is taken vertically between two confinement gates as depicted in Fig.1 A. Here the $t_3$ oxide thickness can be determined as ~14.4 nm, indicating that the etch after the second gate layer nearly completely removed the deposited $t_2$ oxide (vertically, not sideways), which is as expected. The width and length of the plunger gate, and thus also the quantum dot, can be extracted from Fig. A3 A and Fig. A3 D as 50 and 73 nm respectively. These values are within expectations of the designed width and length of 50 and 70 nm.

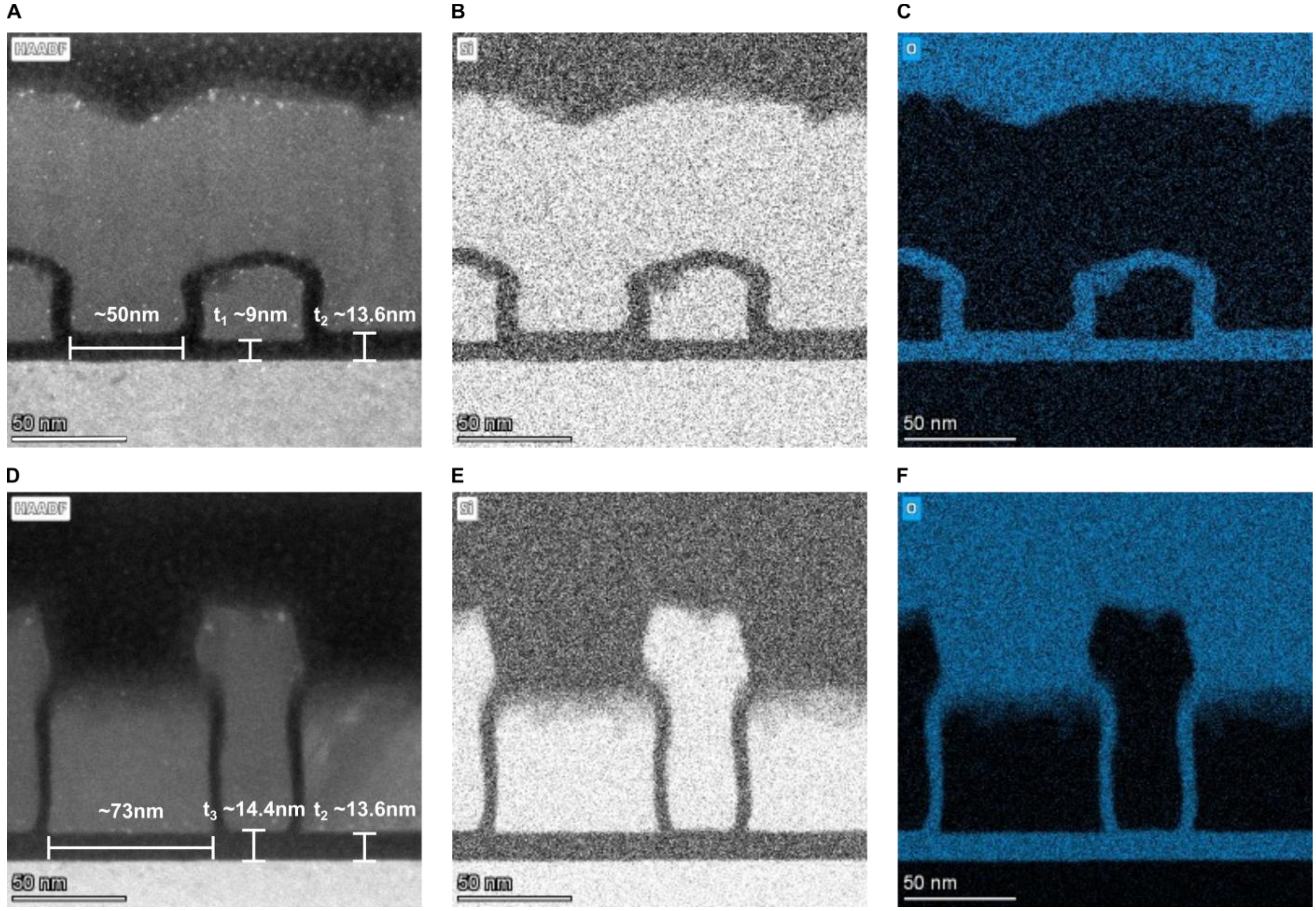

**Fig. A3. TEM and EDS images.** (**A**) HAADF-STEM of a horizontal cut through the array, following a plunger gate. The $t_1$ and $t_2$ oxide thicknesses can be determined as 9 and 13.6 nm respectively, while the plunger width is measured to be 50 nm. (**B**) EDS image of the same cross-section as A, depicting the Si atom distribution. The Si substrate and poly-Si gates are clearly visible. (**C**) EDS image of the same cross-section as A, depicting the O atom distribution. The $SiO_2$ between the substrate and gates is visible in blue. (**D**) HAADF-STEM of a vertical cut through the array, between two confinement gates. The $t_2$ and $t_3$ oxide thicknesses can be determined as 13.6 and 14.4 nm respectively, while the plunger length is measured to be 73 nm. (**E**) EDS image of the same cross-section as D, depicting the Si atom distribution. The Si-substrate and poly-Si gates are clearly visible. (**F**) EDS image of the same cross-section as D, depicting the O atom distribution. The $SiO_2$ between the substrate and gates is visible in blue.

### *Electron temperature*

The electron temperature $T_0$ of the experimental setup was determined by measuring the broadening of the Coulomb oscillations as a function of the lattice temperature. The source-drain bias is chosen sufficiently low such that it has no influence on the Coulomb peak broadening, making the broadening only dependent on the lattice and electron temperatures in the system [46–48]. The shape of the Coulomb peak conductance $G$ will depend on the characteristic energies of the system, the charging energy $E_C$, the level splitting $\Delta_E$ and thermal energy $k_BT$, with $k_B$ the Boltzmann constant and $T$ the temperature.
For a system where $k_BT \ll \Delta_E \ll E_C$, the shape of the Coulomb peaks can be calculated as

$$G = G_{\mathrm{max}} \cosh^{-2}\left(\frac{\Delta_\epsilon}{2k_B T_0}\right) \tag{S1}$$

where $G_{max}$ is a prefactor depending on the tunnel rates in and out of the dot, as well as the temperature of the system, $k_B$ the Boltzmann constant and $\Delta_\epsilon$ the energy difference from the resonance condition.
In this work, the level splitting energy $\Delta_E$ is determined to be ~0.7 meV, while the charging $E_C$ energy is ~4 meV. The thermal energy $k_BT$ is expected to be below 0.26 meV, corresponding to a temperature below 3 K, allowing us to extract the electron temperature using the previously stated formula S1.
The extracted electron temperature $T_0$ is used to calculate the effective electron temperature $T_e$, where the contribution of the phonon temperature $T_{ph}$ and source-drain broadening $T_{sd}$ are considered:

$$T_e = \sqrt{T_0^2 + T_{\mathrm{ph}}^2 + T_{\mathrm{sd}}^2} \tag{S2}$$

For this work, the phonon temperature is taken as the sample temperature, while the source-drain broadening is calculated from the source-drain bias. This results in an electron temperature of 1.4 ± 0.06 K (Fig. A4).

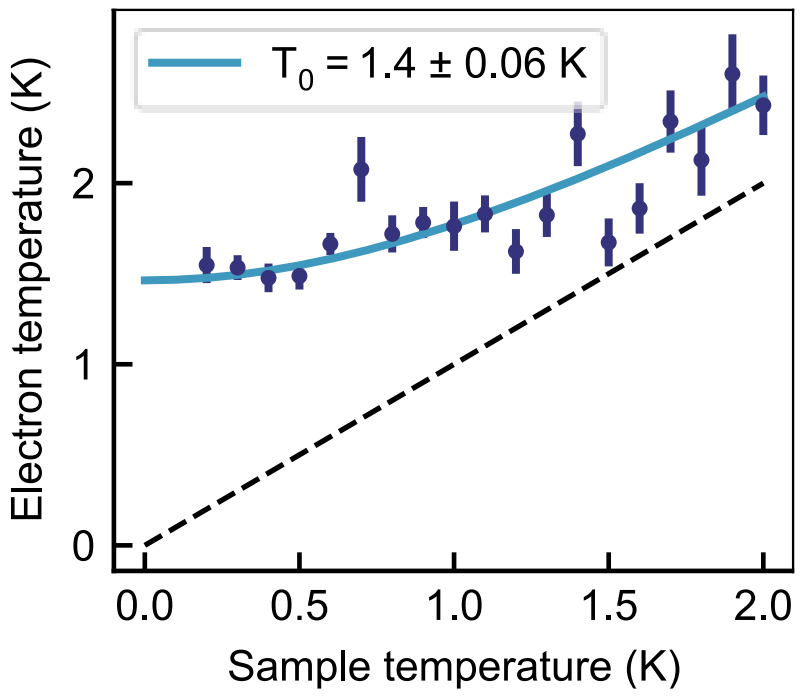

**Fig. A4. Electron temperature extraction.** Error bars on the extracted electron temperatures ($T_0$) of Coulomb oscillations in dot (5,3) from sample H are shown in blue. The measurements were performed at different sample temperatures with a constant source-drain bias of 40 µV. The solid line is a fit of the effective electron temperature ($T_e$) using the extracted electron temperature ($T_0$), phonon temperature ($T_{ph}$) and source-drain broadening ($T_{sd}$) extracted from the measurements. The black dashed line is where the electron temperature equals the fridge temperature and represents the ideal situation.

***All quantum dot data***

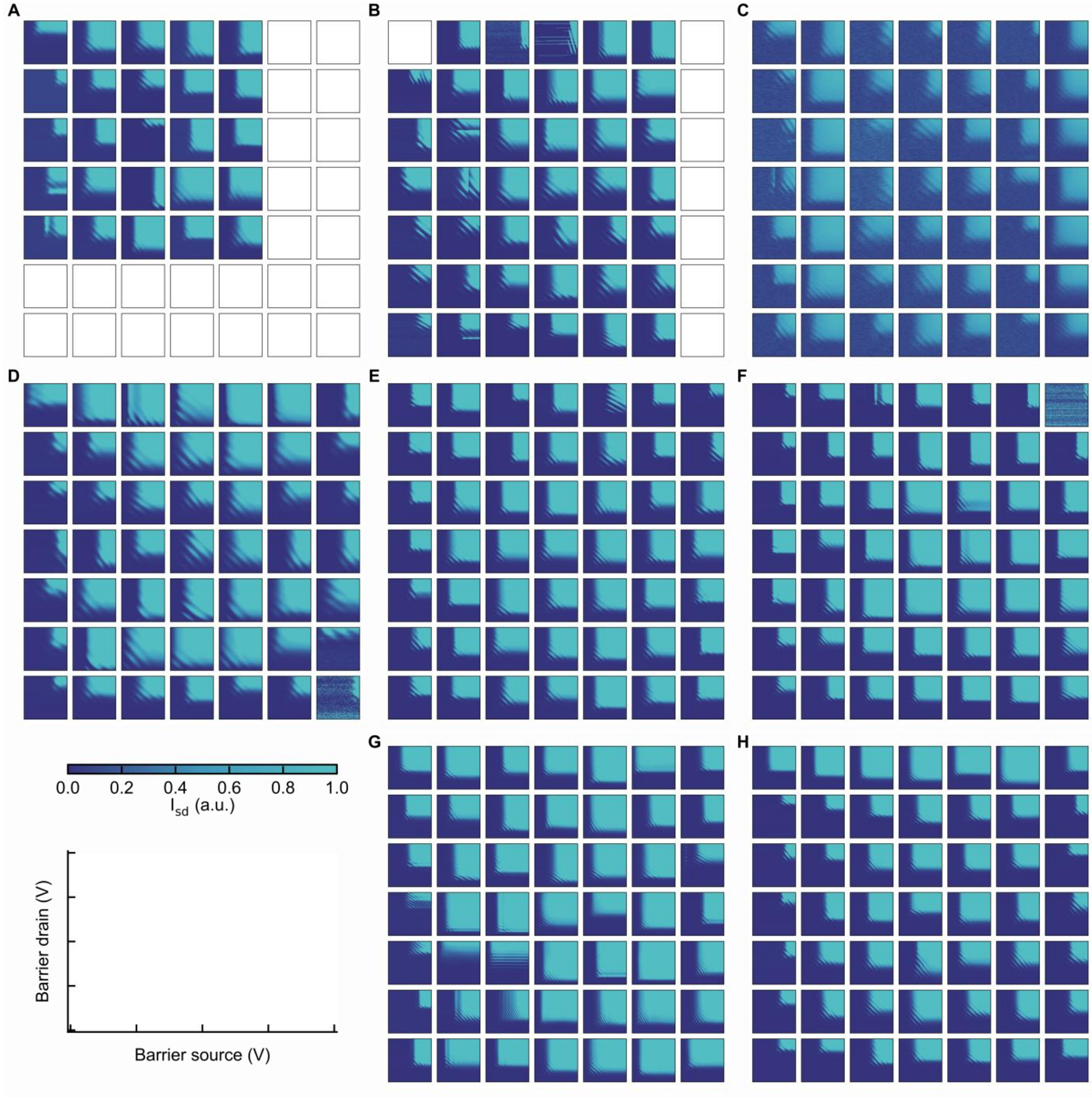

**Fig. A5. All barrier maps.** (**A-H**) Barrier maps for each sample in Table 1, following the same lettering, extracted as in Fig. 2 A. (**A**, **B**) Samples with a $t_1$ oxide thickness of 8 nm. (**C**, **D**)

Samples with a $t_1$ oxide thickness of 12 nm. (**E**, **F**) Samples with a $t_1$ oxide thickness of 15 nm. (**G**, **H**) Samples with a $t_1$ oxide thickness of 20 nm.

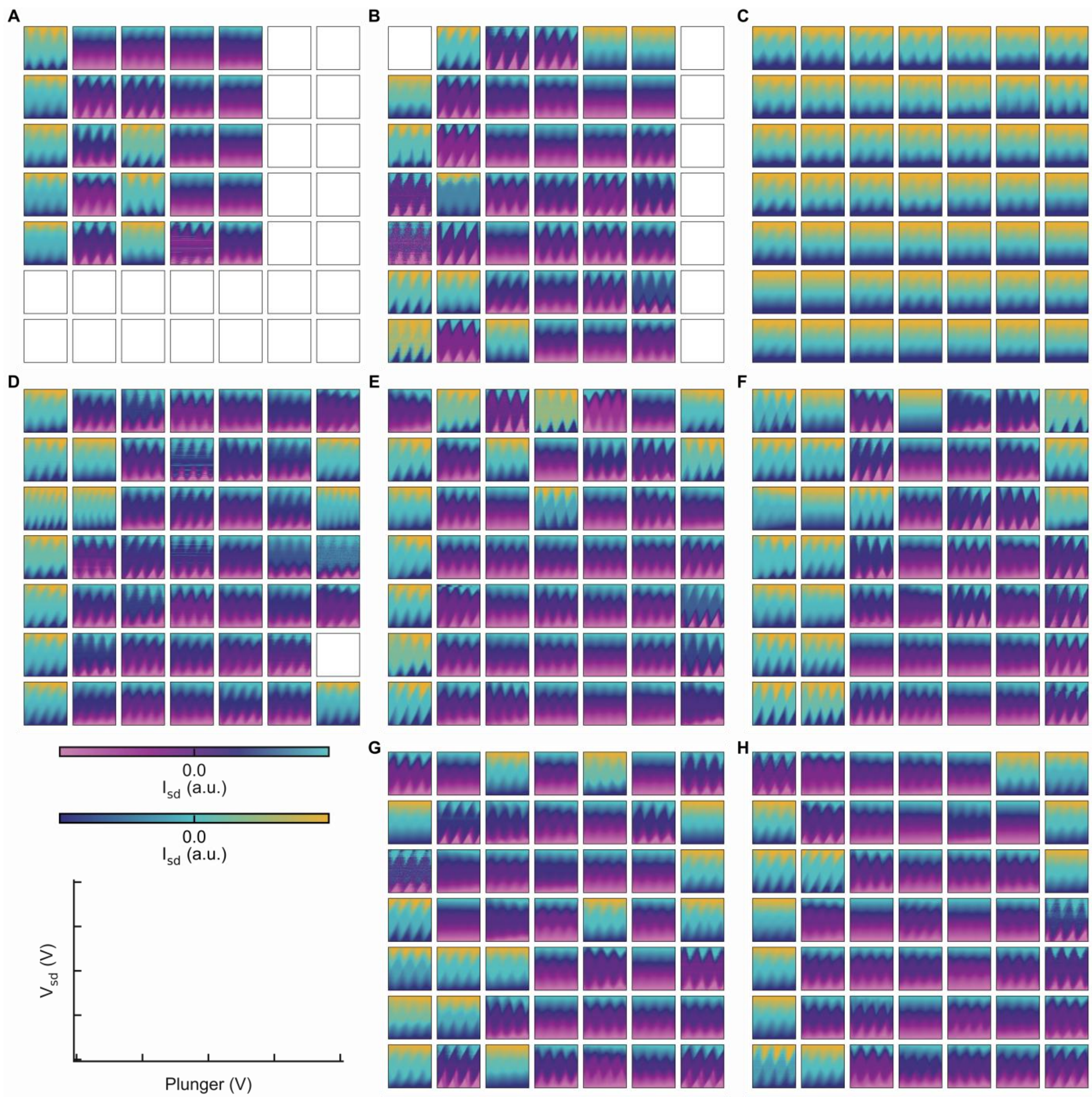

**Fig. A6. All Coulomb diamonds.** (**A-H**) Coulomb diamonds for each sample in Table 1, following the same lettering, extracted as in Fig. 2 B. Shared biased measurements are plotted in blue-purple, individually biased diamonds in yellow-blue, while failed measurements are represented with by an empty plot. (**A**, **B**) Samples with a $t_1$ oxide thickness of 8 nm. (**C**, **D**)

Samples with a $t_1$ oxide thickness of 12 nm. (**E**, **F**) Samples with a $t_1$ oxide thickness of 15 nm. (**G**, **H**) Samples with a $t_1$ oxide thickness of 20 nm.

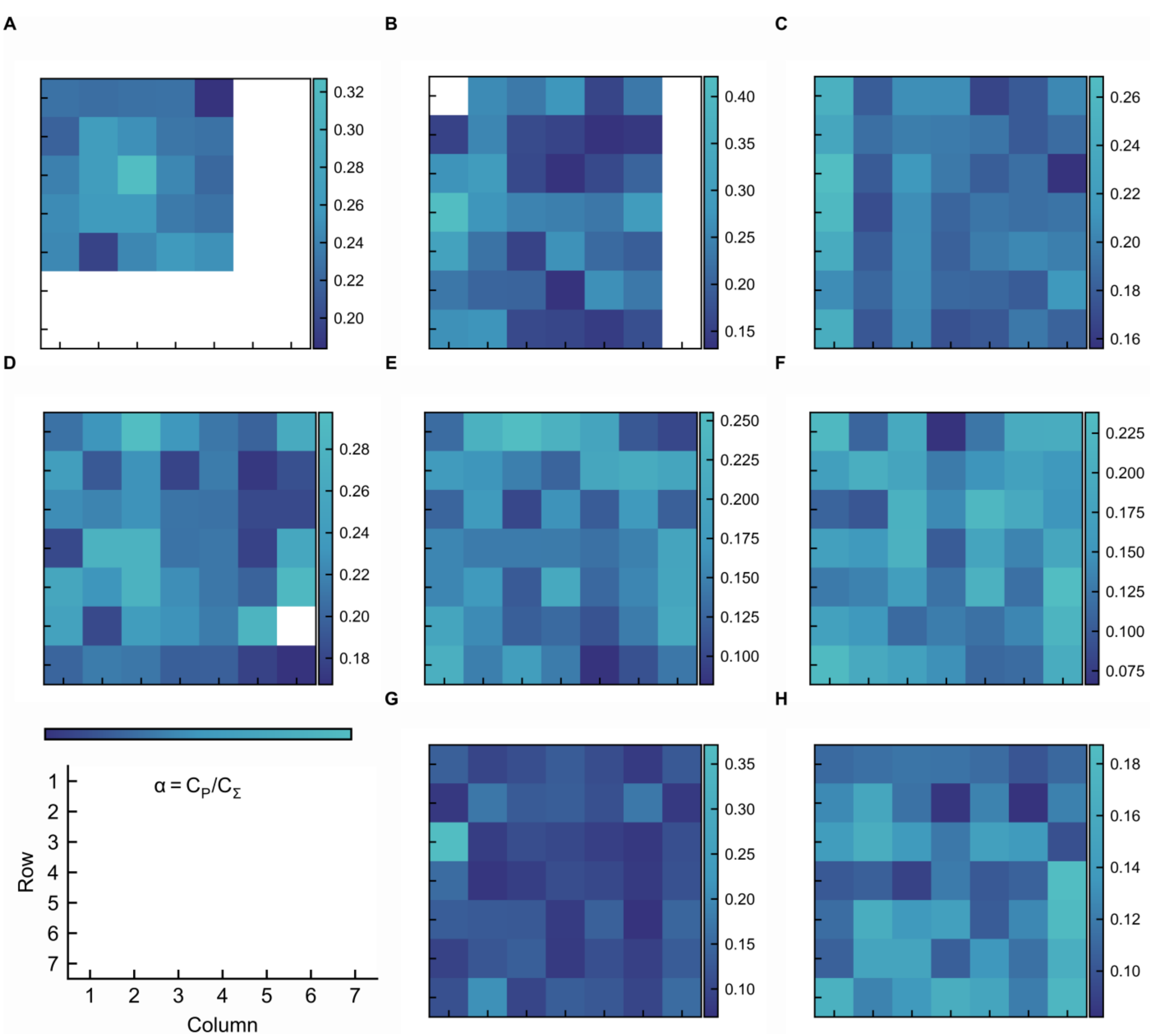

**Fig. A7. All lever arms (α).** (**A-H**) Lever arm (α) for each sample in Table 1, following the same lettering, extracted as in Fig. 3 C. (**A**, **B**) Samples with a $t_1$ oxide thickness of 8 nm. (**C**,

**D**) Samples with a $t_1$ oxide thickness of 12 nm. (**E**, **F**) Samples with a $t_1$ oxide thickness of 15 nm. (**G**, **H**) Samples with a $t_1$ oxide thickness of 20 nm.

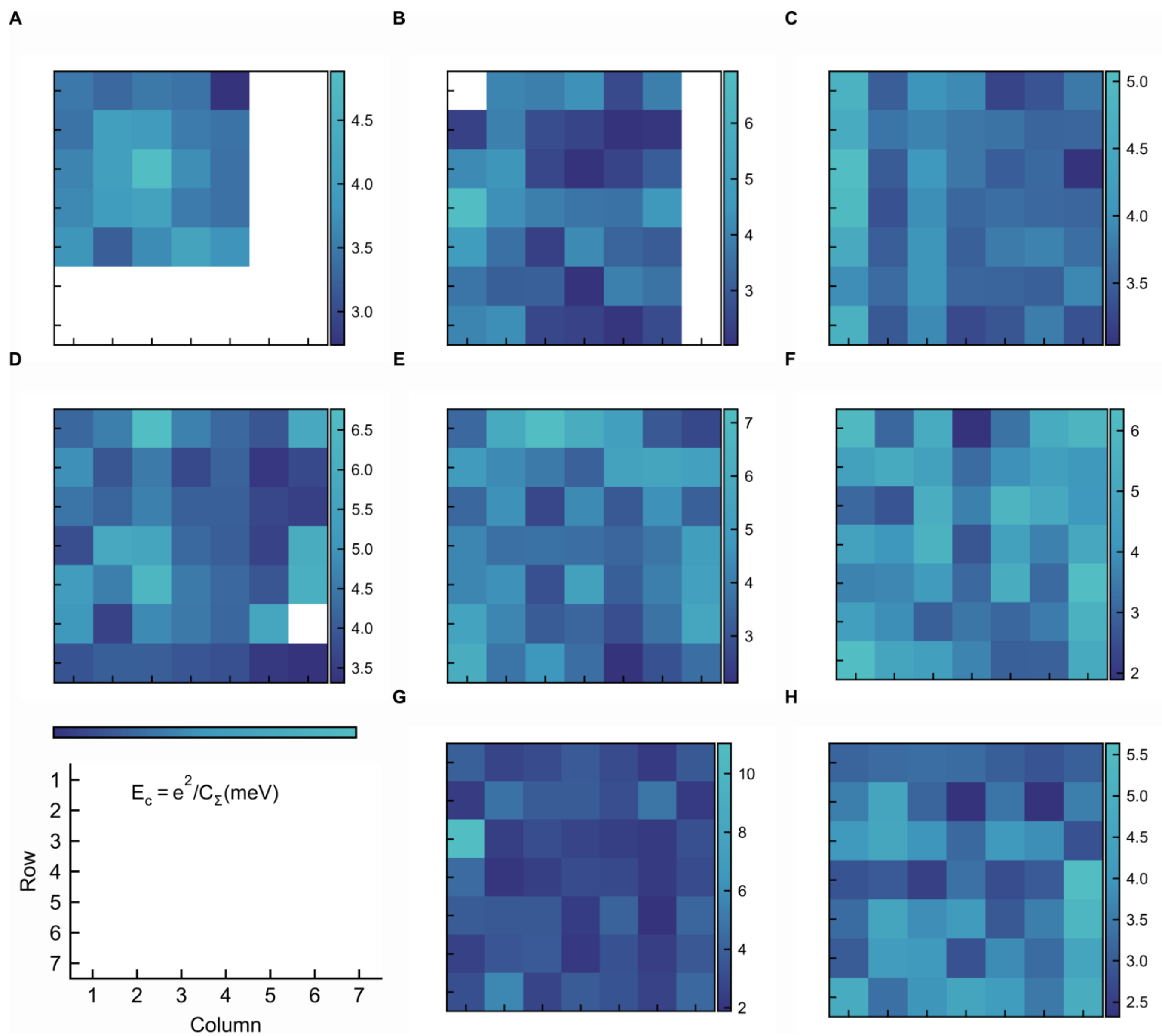

**Fig. A8. All charging energies ($E_C$).** (**A-H**) Charging energy ($E_C$) for each sample in Table 1, following the same lettering, extracted as $E_C = e^2/C_\Sigma$. (**A**, **B**) Samples with a $t_1$ oxide thickness of 8 nm. (**C**, **D**) Samples with a $t_1$ oxide thickness of 12 nm. (**E**, **F**) Samples with a $t_1$ oxide thickness of 15 nm. (**G**, **H**) Samples with a $t_1$ oxide thickness of 20 nm.